\documentclass[journal]{IEEEtran}

\usepackage{amsmath,amsfonts}
\usepackage{amssymb}
\usepackage{amstext}
\usepackage{array}
\usepackage[caption=false,font=normalsize,labelfont=sf,textfont=sf]{subfig}
\usepackage{textcomp}
\usepackage{stfloats}
\usepackage{url}
\usepackage{verbatim}
\usepackage{adjustbox}
\usepackage{cite}
\usepackage{physics}
\usepackage{comment}
\usepackage[ruled,lined]{algorithm2e}

\usepackage{palatino}
\usepackage{tikz}
\usetikzlibrary{graphs, positioning, quotes,shapes.geometric}

\usetikzlibrary{cd}
\usetikzlibrary{quantikz}

\ifCLASSINFOpdf
  \graphicspath{{../}}
  \DeclareGraphicsExtensions{.pdf,.jpeg,.png,.tif}
\else
\fi

\begin{document}

\title{A Variation-Aware Quantum Circuit Mapping Approach Based on Multi-agent Cooperation}
%
%
%

\author{Pengcheng~Zhu,~\IEEEmembership{Student Member,~IEEE,}
        Weiping~Ding,~\IEEEmembership{Senior Member,~IEEE,}
        Lihua~Wei,
        Xueyun~Cheng,~Zhijin~Guan,~\IEEEmembership{Member,~IEEE,}
        and~Shiguang~Feng,~\IEEEmembership{Member,~IEEE}
\thanks{This work was supported by the National Natural Science Foundation of China under Grant 62072259, in part by Jiangsu Province Natural Science Foundation of China under Grant BK20151274, in part by Suqian Science and Technology Foundation under Grant S201819, and in part by the Scientific Innovation Foundation of Suqian University under Grant 2021pt02.}
\thanks{P. Zhu is with the School of Information Science and Technology, Nantong University, Nantong 226019, China, and is with also the Department of Artificial Intelligence, Suqian University, Suqian 223800, China. (e-mail: zhupc@ ntu.edu.cn).}
\thanks{W.ding, X. Cheng, Z. Guan and S. Feng are with the School of Information Science and Technology, Nantong University, Nantong 226019, China (e-mail: dwp9988@163.com; chen.xy@ntu.edu.cn; guan.zj@ntu.edu.cn; shigfeng@ntu.edu.cn). L. Wei is with the Department of Information and Computing Science, Suqian University, Suqian 223800, China. (e-mail: wlh\_sqc@163.com).}
}

\markboth{Journal of \LaTeX\ Class Files,~Vol.~14, No.~8, August~2021}%
{Shell \MakeLowercase{\textit{et al.}}: A Variation-Aware Quantum Circuit Mapping Approach Based on Multi-agent Cooperation}


\maketitle

\begin{abstract}
The quantum circuit mapping approach is an indispensable part of the software stack for the noisy intermediate-scale quantum (NISQ) device. It has a significant impact on the reliability of computational tasks on NISQ devices. To improve the overall fidelity of physical circuits, we propose a quantum circuit mapping method based on multi-agent cooperation. This approach considers the Spatio-temporal variation of quantum operation quality on the NISQ device when inserting ancillary operations. It consists of two core components: the qubit placement algorithm and the qubit routing method. The qubit placement algorithm exploits the iterated local search framework to find a desirable initial mapping for the reduced symmetric form of the original circuit. The qubit routing method generates the physical circuit through multi-agent communication and collaboration. Each agent inserts the ancillary gates independently according to its environment state. The quality of the physical circuit evolves according to an information-exchanging mechanism between agents, which combines the local search and global search. The experimental results confirm the performance of our approach in improving circuit fidelity. Compared with the state-of-the-art method, our method can improve the success rate by 25.86\% on average and 95.42\% at maximum.  
\end{abstract}


\section{Introduction}

\IEEEPARstart{I}{n} the past few years, remarkable progress has been achieved in the physical realization of noisy intermediate-scale quantum (NISQ) devices. Google has released a 53-qubit processor and demonstrated the quantum advantages\cite{arute2019quantum} on it. IBM has also released a 127-qubit processor and will release quantum processors with more than 1,000 qubits in 2023. These NISQ devices have a great prospect in chemistry\cite{tavernelli2020resource}, finance\cite{egger2020credit}, and artificial intelligence\cite{jiang2021co}.
 
With the rapid progress of quantum hardware technology, the software stack for NISQ devices has also attracted more and more attention from both academia and industry. Quantum circuit mapping is an essential part of the kernel software for near-term quantum devices. It converts the quantum logic circuit that is hardware agnostic to the physical one that is hardware compliant by inserting additional quantum operations (such as SWAP gates).
Those inserted operations will increase the gate count and the circuit depth, deteriorating the reliability of computation. Therefore, it is critical to minimize the number of additional gates to ensure the output fidelity of the hardware-compliant circuit. The problem of quantum circuit mapping is NP-complete\cite{siraichi2018qubit, botea2018complexity}, so heuristic approaches\cite{zulehner2017one,kole2019improved,cowtan2019qubit,childs2019circuit,li2019tackling,zhong2020quantum,li2020qubit,zhu2020dynamic,lao2021timing} play a dominant role in efficiently solving this problem. As far as we know, the work in \cite{zulehner2017one} proposes the first quantum circuit mapping method for the real-world quantum device, which divides the quantum circuit into layers and finds a topology compatible solution for each layer using the A* algorithm. The work in \cite{li2020qubit} presents a SWAP-based heuristic search algorithm, which employs the reverse traversal technique to optimize the initial mapping. The work in \cite{cowtan2019qubit} demonstrates the quantum circuit mapping method used in the quantum computing compiler $t\ket{ket}$. The work in ~\cite{li2020qubit} proposes a quantum circuit mapping method based on subgraph isomorphism and depth-limited search. All these heuristic approaches aim to minimize the number of additional gates inserted during the circuit mapping process, oblivious to the variation of operation quality on the actual NISQ devices. However, in the current NISQ devices, the Spatio-temporal variation of qubit and quantum operation quality is inevitable and noticeable.  It is crucial to consider this variation in the circuit mapping approach for the high fidelity of quantum computation. To this end, some recent works ~\cite{murali2019noise,tannu2019not,nishio2020extracting,niu2020hardware} propose the variation-aware approach and demonstrate its effectiveness in improving circuit fidelity. However, their experimental evaluation on only very small-scale circuits (with up to 5 qubits and dozens of gates) is not enough to fully demonstrate the advantages of the variation-aware strategy in improving fidelity. Furthermore, these approaches are essentially circuit mapping processes based on a single intelligent agent. Since these single-agent approaches generally use the greedy search strategy, the solutions obtained usually have considerable space for further optimization~\cite{tan2020optimality} in both gate count and circuit fidelity. 

As discussed, although some effort has been dedicated to the variation-aware quantum circuit mapping approach, the noted algorithms suffer from the following limitations and challenges: (1) The single-agent scheme based on a simple accumulation of the local optima leaves considerable space for further optimization. (2) It is not comprehensive enough to perform experiments on only small-scale circuits.

To address these challenges, we propose a novel quantum circuit mapping scheme based on multi-agent cooperation.  This approach considers the variation of gate errors and aims to optimize the overall reliability of quantum computation on actual NISQ devices. Our main contributions include: 

\begin{enumerate}
	\item We propose a qubit placement algorithm based on the iterated local search for the reduced symmetric form of the original logical circuit. This algorithm can guarantee both the quality and the diversity of the initial population of agents.
	
	\item We also propose a qubit routing algorithm based on multi-agent cooperation. Each agent makes decisions independently according to the environmental status and evolves the circuit quality through a mechanism similar to the shuffled frog-leaping algorithm (SFLA). 
		
	\item We perform the experiments on a noisy simulator with gate error 10x lower than that of the state-of-the-art device of IBM. With the help of this simulator, we can reliably execute large-scale circuits beyond the capacity of current NISQ devices. Extensive experimental evaluation on various benchmarks confirms the effectiveness of our approach in improving circuit fidelity.
\end{enumerate}

This paper is organized as follows. We briefly review the NISQ device and quantum circuit mapping in Section~\ref{sec-backgd}. Section~\ref{sec-algo-vitual} and Section~\ref{sec-multi-agent} delineate the qubit placement algorithm and the qubit routing algorithm, respectively.  Section~\ref{sec-experiment} shows the experimental evaluation. Section~\ref{sec-discussion} gives further discussions and Section~\ref{sec-conclusion} concludes this paper.

\section{Background}\label{sec-backgd}

This section will review the characteristics of current NISQ devices through several quantum devices of IBM and briefly introduce the quantum circuit mapping problem.

\subsection{The Characteristics of NISQ devices}

\begin{figure}[!t]
	\centering
	\begin{adjustbox}{width=0.4\textwidth}
	\subfloat[IBMQ\_belem]{
		\begin{tikzcd}[every arrow/.append style={dash}]
			\tikz{\node[draw, circle, inner sep=2pt]{0}}  \arrow[r]  & \tikz{\node[draw, circle, inner sep=2pt]{1}}  \arrow[r] \arrow[d] & \tikz{\node[draw, circle, inner sep=2pt]{2}}  \\
			& \tikz{\node[draw, circle, inner sep=2pt]{3}}  \arrow[d] &  \\
			& \tikz{\node[draw, circle, inner sep=2pt]{4}}	& 
		\end{tikzcd}
		\label{fig-quanarcht-a}}
	\hfil
	\subfloat[IBMQ\_guadalupe]{
			\begin{tikzcd}[every arrow/.append style={dash}]
			&  &  & \tikz{\node[draw, circle, inner sep=2pt]{6}}  \arrow[d] &  &  &  \\ 
			\tikz{\node[draw, circle, inner sep=2pt]{0}}  \arrow[r]	& \tikz{\node[draw, circle, inner sep=2pt]{1}}  \arrow[r] \arrow[d] & \tikz{\node[draw, circle, inner sep=2pt]{4}}  \arrow[r]
			& \tikz{\node[draw, circle, inner sep=2pt]{7}}  \arrow[r] & \tikz{\node[draw, circle, inner sep=1pt]{10}}  \arrow[r] &  \tikz{\node[draw, circle, inner sep=1pt]{12}}  \arrow[r] \arrow[d] 
			& \tikz{\node[draw, circle, inner sep=1pt]{15}}  \\
			& \tikz{\node[draw, circle, inner sep=2pt]{2}}  \arrow[d] &  &  &  & \tikz{\node[draw, circle, inner sep=1pt]{13}}   \arrow[d] &  \\
			& \tikz{\node[draw, circle, inner sep=2pt]{3}}  \arrow[r] & \tikz{\node[draw, circle, inner sep=2pt]{5}}  \arrow[r] & \tikz{\node[draw, circle, inner sep=2pt]{8}}  \arrow[r]  \arrow[d] 
			& \tikz{\node[draw, circle, inner sep=1pt]{11}}  \arrow[r] &  \tikz{\node[draw, circle, inner sep=1pt]{14}}   & \\
			&  &  & \tikz{\node[draw, circle, inner sep=2pt]{9}}   &  &   & 
		\end{tikzcd}
		\label{fig-quanarcht-b}}
	\end{adjustbox}
	\caption{The coupling graphs of two IBM's quantum devices.}
	\label{fig-quanarcht}
\end{figure}
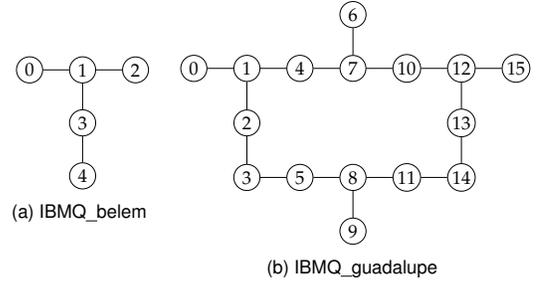

IBM's quantum devices are designed based on the superconducting transmon qubit. Such a qubit of this kind is not a natural atom, but an artificial circuit made of superconducting materials. For superconducting qubits, the single-qubit gates are usually driven resonantly by a microwave pulse, and the two-qubit gates can only act on the two qubits connected by a coplanar waveguide (CPW) resonator.  Fig. \ref{fig-quanarcht} shows the hardware topology (also called coupling graph) of two IBM's quantum devices. Each node in Fig. \ref{fig-quanarcht} represents an individual qubit, and each edge represents a CPW bus. In the rest of the paper, we denote the coupling graph by $CG(V,E)$, where $V$ and $E$ represent the node-set and the edge-set, respectively. Accordingly, $|V|$ and $|E|$ represent the cardinality of the set $V$ and the set $E$, respectively. In particular, we refer to the qubits in the hardware topology as physical qubits.

\begin{figure}[!t]
	\centering
	\includegraphics[scale=0.3]{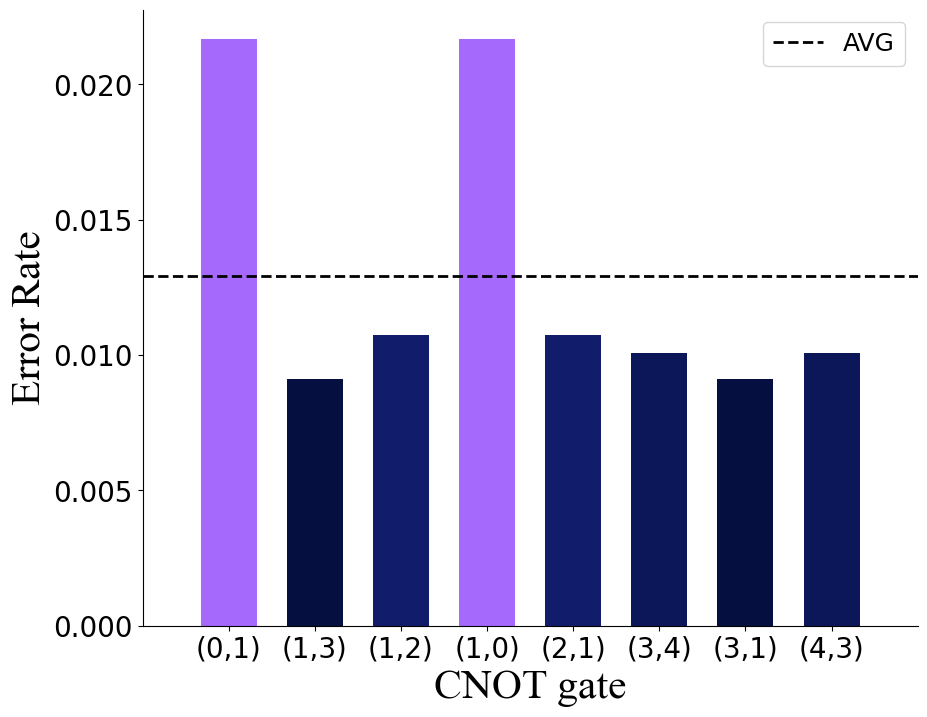}
	\caption{The distribution of CNOT gate error rates}
	\label{fig-cxerror}
\end{figure}

Due to the defects of manufacturing technology and the error of the control pulse signal, the superconducting qubits and quantum gates acting on them show remarkable variability in quality. Even the same quantum gate behaves differently on different qubits. Fig. \ref{fig-cxerror} shows the distribution of CNOT gate errors on the 5-qubit NISQ device IBMQ\_belem (Fig. \ref{fig-quanarcht-a}). We can easily observe that the CNOT gate error varies apparently as the qubits change. Furthermore, the qubit quality and operation fidelity will inevitably drift over time, so IBM calibrates their devices periodically, at least once a day, to ensure everything is under control.

At present, the NISQ devices suffer from many noise sources, such as gate error, decoherence error, and readout error. In this paper, we focus on the variation of gate error, especially on that of two-qubit gate error, since the two-qubit gate error~\cite{murali2019noise,tannu2019not,nishio2020extracting} is a major source of noise on NISQ devices. 

\subsection{Quantum circuit mapping}

The quantum circuit realizing a specific quantum algorithm is usually hardware-agnostic. We call it the quantum logical circuit to distinguish it from the physical circuit directly runnable on real quantum devices. We denote the quantum logical circuit as $LC(Q,G)$, where $Q$ and $G$ represent the set of logical qubits and the set of quantum gates, respectively.  

The coupling graph $CG(V,E)$ of a NISQ device imposes a topology constraint on physically executing the quantum circuit. This topology constraint requires that every two-qubit gate should act on the two qubits connected by a bus. The quantum logical circuit with a sequence of two-qubit gates generally cannot satisfy this constraint. Quantum circuit mapping can make the quantum circuit compatible with the restricted topology through two main steps. The first step is the qubit placement, and the second is the qubit routing.

The qubit placement is to place each logical qubit of the circuit onto a physical qubit of the device.  Mathematically, the relationship between logical and physical qubits is an injective function $\pi$ from the logical qubits $Q$ to the physical qubits $V$. We call this mapping relationship the initial mapping. 

\begin{figure}[!t]
	\centering
	\begin{adjustbox}{width=0.25\textwidth}
	\begin{quantikz}
	 &  \swap{1} & \qw \\
     &  \targX{}  & \qw
	\end{quantikz}
	$\equiv$
	\begin{quantikz}
		& \ctrl{1} & \targ{}    & \ctrl{1} & \qw \\
		& \targ{}  & \ctrl{-1} & \targ{}  & \qw
	\end{quantikz}
	\end{adjustbox}
	\caption{The decomposition of a SWAP gate.}
	\label{fig-decop-swap}
\end{figure}
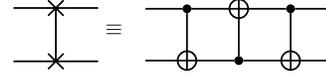

Given an initial mapping, the qubit routing moves all the two-qubit gates that do not meet the topology constraint to the two adjacent nodes in the coupling graph by inserting SWAP gates. The SWAP gate is a quantum operation that can interchange the quantum states of two qubits, and it can be decomposed into 3 CNOT gates, as shown in Fig. \ref{fig-decop-swap}. In the qubit routing process, we insert a series of SWAP gates to move the logical qubits along the edge of the coupling graph until all two-qubit gates act on adjacent nodes. The circuit obtained after routing satisfies the topology constraint and thus is physically executable. We call it the physical circuit and denote it as  $PC(V, G)$, where $V$ and $G$ are physical qubit set and gate set, respectively.

We give an example of the quantum circuit mapping problem as follows. For more illustrative examples, please refer to the previous work~\cite{zulehner2017one,li2019tackling,li2020qubit,zhu2020dynamic}. 

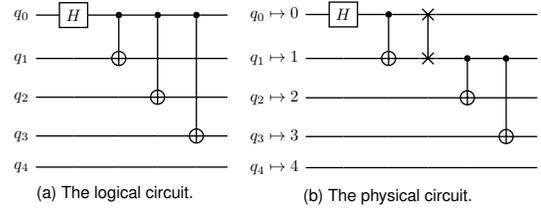
\begin{figure}[!t]
	\centering
	\begin{adjustbox}{width=0.4\textwidth}
	\subfloat[The logical circuit.]{
		\begin{quantikz}
		\lstick{$q_0$} & \gate{H} & \ctrl{1} & \ctrl{2} & \ctrl{3} & \qw \\
		\lstick{$q_1$} & \qw      & \targ{}  & \qw	    & \qw	   & \qw \\
		\lstick{$q_2$} & \qw      & \qw	     & \targ{}  & \qw	   & \qw \\
		\lstick{$q_3$} & \qw      & \qw	     & \qw	    & \targ{}  & \qw \\
		\lstick{$q_4$} & \qw      & \qw	     & \qw	    & \qw	   & \qw
	\end{quantikz}
		\label{fig-GHZtoIBM-a}}

	\subfloat[The physical circuit.]{
		\begin{quantikz}
		\lstick{$q_0\mapsto 0$} & \gate{H}  & \ctrl{1} & \swap{1} 	 & \qw      & \qw 	   & \qw \\
		\lstick{$q_1\mapsto 1$} & \qw		& \targ{} & \targX{} 	 & \ctrl{1}	& \ctrl{2} & \qw \\
		\lstick{$q_2\mapsto 2$} & \qw		& \qw      & \qw	     & \targ{}  & \qw	   & \qw \\
		\lstick{$q_3\mapsto 3$} & \qw		& \qw      & \qw	     & \qw	    & \targ{}  & \qw \\
		\lstick{$q_4\mapsto 4$} & \qw		& \qw      & \qw	     & \qw	    & \qw	   & \qw
		\end{quantikz}
		\label{fig-GHZtoIBM-b}}
	\end{adjustbox}
	\caption{Mapping GHZ to IBMQ\_belem.}
	\label{fig-GHZtoIBM}
\end{figure}

\textit{Examaple 1}. We show a quantum logical circuit for preparing the GHZ state in Fig. \ref{fig-GHZtoIBM-a} and map it to IBMQ\_belem (Fig. \ref{fig-quanarcht-a}). Given the identity initial mapping between the logical qubits and the physical qubits, we insert a SWAP gate to get the physical circuit (Fig. \ref{fig-GHZtoIBM-b}), which is topology compliant.   

When performing circuit mapping tasks, considering the variation of operation quality is  crucial to improve computational reliability. Different strategies for qubit placement and qubit routing will lead to physical circuits with different fidelity. Recent work~\cite{murali2019noise,tannu2019not,nishio2020extracting} provides some motivating examples for the importance of considering error variation.

\section{Qubit placement algorithm}\label{sec-algo-vitual}

Our multi-agent qubit routing approach requires numerous initial mappings, one for each agent. The quality and diversity of the initial agent population largely depend on the initial mappings generated by the qubit placement algorithm. However, most existing qubit placement algorithms~\cite{li2020qubit, zhu2020dynamic} are deterministic and thus lack diversity for the resulting initial mapping. The others based on the meta-heuristic algorithm, such as simulated annealing~\cite{zhou2020quantum, niu2020hardware}, are stochastic but can not guarantee the quality of the generated initial mapping within a reasonable time. Therefore, we propose a qubit placement algorithm based on the iterated local search (ILS) to ensure both the quality and diversity of the initial mapping. In this section, we first introduce several related definitions, then present the construction of the reduced symmetric circuit, and finally focus on the ILS-based qubit placement method.

\subsection{Related concepts}

\emph{Definition 1:} For a CNOT gate $CX(q_i,q_j)$, its physical distance on the coupling graph $CG(V,E)$ under the initial mapping $\pi$ is the length of the shortest path between the two physical nodes $\pi(q_i)$ and $\pi(q_j)$ on $CG$.

\emph{Definition 2:} The front layer of a quantum logical circuit refers to the gates without any dependence on other gates.

The dependence between gates in the logcial circuit can be expressed as a directed acyclic graph (DAG)~\cite{li2019tackling, childs2019circuit}. In the DAG, all the nodes without any predecessors constitute the front layer, that is, the first layer of the circuit. By removing the nodes in the front layer, we can derive the second layer of the circuit. In this way, we can partition the logical circuit into multiple layers from left to right.

\emph{Definition 3:} A two-qubit quantum gate is executable if it is in the front layer and its physical distance is equal to 1. A single-qubit gate is executable if it is in the front layer.

A two-qubit gate satisfies the topology constraint imposed by the coupling graph only if its physical distance is 1.

\subsection{Reduced symmetric circuit}

We take the reduced symmetric circuit as the input of our ILS-based qubit placement approach. This reduced circuit is a reduced form of the original logical circuit. It preserves the qubit-interaction structure in the original circuit while retaining only a small number of gates of the original circuit.  To create a reduced symmetric circuit, we scan the original circuit from left to right, skipping each single-qubit gate, and remaining only the first two-qubit gate for each interacting qubit pair. The reduced circuit plus its mirror circuit constitutes the reduced symmetric circuit. The mirror circuit is obtained by reversing the gate order of the reduced circuit. 

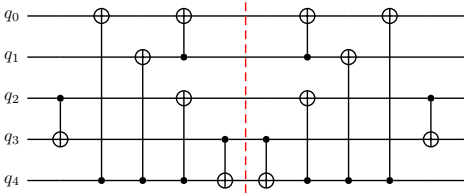
\begin{figure}[!t]
	\centering
	\begin{adjustbox}{width=0.35\textwidth}
	\begin{quantikz}
		\lstick{$q_0$} & \qw        & \targ{}   & \qw        & \targ{}    & \qw\slice{} & \qw	   	& \targ{}   & \qw       & \targ{}  	& \qw      & \qw \\
		\lstick{$q_1$} & \qw        & \qw       & \targ{}    & \ctrl{-1}  & \qw         & \qw 		& \ctrl{-1} & \targ{}   & \qw      	& \qw      & \qw \\
		\lstick{$q_2$} & \ctrl{1}   & \qw	    & \qw        & \targ{}	  & \qw         & \qw       & \targ{}   & \qw       & \qw	 	& \ctrl{1} & \qw \\
		\lstick{$q_3$} & \targ{}    & \qw	    & \qw	     & \qw        & \ctrl{1}    & \ctrl{1}  & \qw       & \qw       & \qw  		& \targ{}  & \qw \\
		\lstick{$q_4$} & \qw        & \ctrl{-4}	& \ctrl{-3}	 & \ctrl{-2}  & \targ{}     & \targ{}   & \ctrl{-2} & \ctrl{-3} & \ctrl{-4} & \qw  	   & \qw
	\end{quantikz}
	\end{adjustbox}
	\caption{Reduced symmetric circuit of '4gt13\_92'}
	\label{fig-rmc}
\end{figure}

We show the reduced symmetric circuit of the benchmark circuit '4gt13\_92' in Fig. \ref{fig-rmc}. The left part of the quantum circuit in Fig. \ref{fig-rmc} is the reduced circuit. There are six different interacting qubit pairs in '4gt13\_92', and we retain only the first two-qubit gate for each interacting qubit pair, so there are six CNOT gates in the reduced circuit. These CNOT gates maintain their dependency on each other in the original circuit. The right part of the quantum circuit is the mirror circuit, which is just the reverse of the reduced circuit.

For any quantum circuit, its reduced symmetric circuit contains a limited number of gates, with an upper bound of twice the number of edges in the coupling graph. The small number of gates ensures that the time overhead of mapping or simulating the reduced circuit is trivial.

\subsection{ILS-based qubit placement algorithm}

Our qubit placement algorithm provides an initial mapping for each agent of the routing algorithm. To ensure the diversity and quality of the initial agent population, it should generate multiple different good initial mappings. An advisable initial mapping allocates the logical qubits to a cluster of adjacent physical qubits with a similar qubit-interaction structure. Moreover, to improve the fidelity of the output state, these selected physical qubits should have long coherence times and strong links with low error rates. The qubit routing process can gather the logical qubits into a cluster of adjacent physical qubits by inserting SWAP gates, thus generating an appropriate initial mapping. With this observation, previous work~\cite{li2019tackling} proposes the bidirectional traversal technique that improves a randomly generated initial mapping by invoking qubit routing forward and backward. Although this approach can generate multiple initial mappings by inputting distinct random initial mappings, the quality of the improved initial mapping largely depends on the random initial mapping and thus is very unstable. Moreover, this approach does not consider the error variation of different qubits and links. 

To better serve our multi-agent approach of qubit routing, we propose a qubit placement algorithm based on the iterated local search~\cite{lourencco2019iterated}. Our qubit placement approach regards the qubit routing process as a discrete function with the initial mapping as input and the physical circuit as output. It searches the domain of this function for the initial mapping that can lead to the physical circuit with high fidelity by the iterated local search. During each local search, the neighborhood of the current initial mapping is generated by qubit routing and evaluated by the fidelity of the resulting physical circuit. The fidelity of a physical circuit is estimated by the rate of successful trials obtained by executing this circuit many times on a noisy simulator that mimics the target devices. It is very time-consuming to route and simulate the entire quantum circuit in an iterated way, especially for the large-scale circuit. In contrast, it is trivial in time overhead for the reduced symmetric circuit.  As discussed above, the reduced circuit can preserve the same qubit-interaction relationship as the original circuit. In addition, its mirror part can guarantee that the updated initial mapping generated by qubit routing is beneficial to the routing of the leftmost part of the original circuit. Therefore, by replacing the original circuit with its reduced symmetric form, we can achieve a good trade-off between the quality of the initial mapping and the time overhead of generating and evaluating the initial mapping.

We show our ILS-based qubit placement approach in Algorithm~\ref{algo1}. Line 1 of Algorithm~\ref{algo1} initializes the initial mapping and the maximum value of estimated fidelity of the resulting physical circuits. Line 4-10 constitute the local search of the initial mapping. With the reduced circuit ($RC$) and the current initial mapping ($\pi_0$) as input, Line 5 invokes the qubit routing approach to generate the physical circuit ($pc_0$) corresponding to $\pi_0$ along with a novel initial mapping ($\pi_1$) for the next iteration. Line 6 calculate the estimated fidelity ($p_{suc}$, percentage of successful trials) by running $pc_0$ multiple times on a noisy simulator. Line 7-9 preserve the best initial mapping leading to the physical circuit with the best fidelity so far. Line 10 prepares for the next local search iteration. After a limited number (defined by $J$) of consecutive local search iterations, the logical qubits are gathered in a local region of the hardware topology by the qubit routing approach. In this case, it is very difficult for the qubit routing approach to move the logical qubits further to the other area of the hardware topology. Therefore, to explore more areas of the hardware topology, we perform a random shuffle perturbation on the best initial mapping so far (Line 3). We repeat this search process $I$ times (Line 2) and return the best initial mapping found (Line 13). We use the qubit routing approach of our previous work (Algorithm 3 of~\cite{zhu2021iterated}) as the qubit router ($qubit\_router()$, Line 5) in Algorithm~\ref{algo1}. However, indeed any heuristic qubit mapping approach based on SWAP gates is adequate. The qubit router of this kind traverses the quantum circuit from left to right and inserts the SWAP gate step by step to make all the two-qubit gates comply with the topology constraint. For more details of such heuristic qubit routers, please refer to the prior work~\cite{li2019tackling,li2020qubit,zhu2020dynamic}.

\LinesNumbered
\begin{algorithm}[!t]
	\KwIn{Reduced symmetric circuit $RC(Q,G)$ and coupling graph $CG(V,E)$}
	\KwOut{Initial mapping $\pi$}
	$(\pi, p\_best)=(rand\_map(),0)$\;
	\While{$i++ \leq I$} {
		$\pi_0 = shuffling\_perturb(\pi)$\;
		\While{$j++ \leq J$}{
			$(pc_0,\pi_1) = qubit\_router(RC,CM,\pi_0)$\;
			$p\_suc = simulate(pc_0,\pi_0)$\;
			\If{$p\_suc > p\_best$}{$(\pi,p\_best) = (\pi_0, p\_suc)$}
			$\pi_0 = \pi_1$\;
		}
	}
	\Return $\pi$\;
	\caption{Qubit placement based on iterated local search}
	\label{algo1}
\end{algorithm}

We also take advantage of the iterated local search in another work of quantum circuit mapping. The ILS-based framework in this paper distinguishes itself from our previous work in three aspects. First and foremost, it searches and outputs the initial mapping producing the best circuit fidelity rather than the physical circuit with minimum additional gates. Second, it accounts for the variation of qubit quality and gate error by running physical circuits on a simulator that mimics the real quantum hardware. Third, it traverses the reduced circuit instead of the original one for time efficiency.   

\subsection{Time complexity}

For the qubit router ($qubit\_router()$), the worst time complexity is $O(|V|^3 \cdotp D \cdotp |G|)$, where $D$ is the diameter of the coupling graph, $V$ is the set of physical qubits, and $G$ is the set of gates in the reduced circuit. There are at most $2|E|$ gates in the reduced circuit. In addition, $|E| \leq |V| \cdotp (|V|-1) / 2 $ and $D < |V|$. Therefore, a rough upper bound of this complexity is $O(|V|^6)$. Accordingly, the worst time complexity of Algorithm 1 is $O(I \cdotp J \cdotp |V|^6)$, which is polynomial.

\section{Qubit routing approach based on multi-agent cooperation}\label{sec-multi-agent}

In this section, we present the multi-agent qubit routing approach. In particular, we emphasize the decision-making of any individual agent and the cooperation mechanism between agents.

\subsection{Motivation}

Most existing approaches for qubit routing aim to minimize the number of additional gates inserted to make the quantum circuit conforms to the hardware topology. Although gate count is an essential metric of the quality of quantum circuits, it is not comprehensive enough. In contrast, the computational success rate of quantum circuits (circuit fidelity) on quantum devices is more comprehensive, especially for NISQ devices. Some previous works of qubit routing take this metric as their first optimization goal. They exploit the variation of gate error to improve the overall circuit fidelity. Such approaches are more practical for nowaday quantum devices because the error variation is common at present. However, existing variation-aware qubit routing approaches insert SWAP gates by local-optimum strategies. Simple accumulation of the local optima generally cannot approach the global optima, so there is still a big gap between the approximate solution obtained by existing approaches and the global optima. 

\begin{figure}[!t]
 \begin{center}
  \begin{tikzpicture}[node distance=10pt]
   \tikzstyle{every node}=[font=\footnotesize]
   \tikzstyle{io} = [trapezium, trapezium left angle=70, trapezium right angle=110]
   
   \node[draw, rounded corners]                        (start)   {Start};
   \node[draw, io, align=center, text width= 2.5cm, below=of start]   (input)  {$S1$: Initialize all agents.};
   \node[draw, diamond, aspect=2, below=of input]      (choice)  {Is done?};
   \node[right=of choice, yshift=0.2cm]  (choight)  {Yes};
   \node[draw, align=center, text width= 3cm, below=of choice]   (step 1)  {$S2$: Rank all agents.};
   \node[draw, align=center, text width= 3cm, below=of step 1]   (step 2)  {$S3$: Partition agents.};
   \node[draw, align=center, text width= 3cm, below=of step 2]   (step 3)  {$S4$: Evolve each agent partition.};
   \node[draw, align=center, text width= 3cm, below=of step 3]   (step 4)  {$S5$: Each agent responds to its environment.};
   \node[draw, align=center, text width= 3cm, below=of step 4]   (output)  {$S6$: Return the physical circuit of the best agent.};
   \node[draw, rounded corners, below=of output]       (end)     {End};

   \graph{
    (start) -> (input) -> (choice) -> ["No"] (step 1) -> (step 2) -> (step 3) -> (step 4);
    (output) -> (end);
   };
  
   \draw [->] ($(choice) + (1.1cm,0)$) -- ++(1.25,0) |- ($(output) + (1.65cm,0)$);
   \draw [->] ($(step 4) + (-1.6cm,0)$) -- ++(-1,0) |- ($(choice) + (-1.05cm,0)$);
  \end{tikzpicture}
 \end{center}
 \caption{\label{fig-flow} Schematic of the qubit routing approach.}
\end{figure}
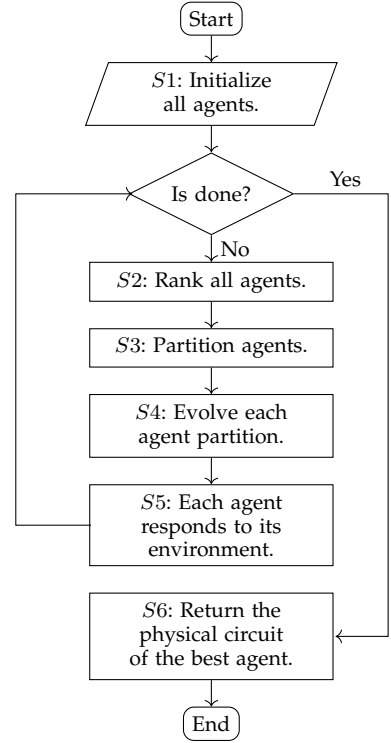

The research on swarm intelligence~\cite{bansal2019evolutionary} shows that information sharing can effectively promote the evolution of the population. With the help of a proper information sharing mechanism, we can also continuously improve the quality of physical circuits for a population of multiple agents. Motivated by this, we present a qubit routing approach based on multi-agent communication and collaboration. We show the schematic of this approach in Fig.~\ref{fig-flow}. As shown, multiple agents complete the quantum circuit mapping task cooperatively. Each agent starts from an initial environment state (see $S1$) and iteratively responds to its environment through a specially selected action (see $S5$) to gradually build the physical circuit. In each iteration, each agent can improve its partial physical circuit through global information exchange (see $S2$) and local information exchange (see $S4$). When the iteration satisfies the termination condition, we take the physical circuit of the best agent as output (see $S6$).

\subsection{Decision-making of individual agent}
An agent perceives its environment and makes decisions about actions to take. For the qubit routing problem,  the status of the environment can be described by three elements. They are the injective mapping relationship ($\pi$) between logical qubits and physical qubits, the remaining logical circuit ($vcir$) to be mapped, and the physical circuit ($pcir$) composed of already mapped gates and inserted SWAP operations. We thus use the triple $(\pi,vcir,pcir)$ to represent the environment state. At the beginning of qubit routing, the environment status is initialized as follows: $\pi$ is an initial mapping returned by Algorithm~\ref{algo1}; $vcir$ is obtained by removing all single-qubit gates and two-qubit gates executable on $\pi$ from the original logical circuit; $pcir$ consists of all executed gates removed from the original logical circuit. To achieve the task of qubit routing, the agent should take action by inserting SWAP gates until the $vcir$ is empty. If there are still gates left in $vcir$, it will decide which SWAP gate to insert according to the following steps.

\emph{Step 1.} Evaluate the reward function for each candidate SWAP gate. Among all possible SWAP gates, those that reduce the physical distance for at least one front gate constitute the candidate SWAP set.  We show the reward function of the SWAP gate ($sw$) in (\ref{equ-reward}), where $\pi$ is the mapping relation between logical and physical qubits, $L_0$ is the set of front gates, $d(g,\pi)$ is the physical distance of the gate $g$. (\ref{equ-reward}) accumulates the reduction in physical distance for all front gates. The larger the value of (\ref{equ-reward}), the higher the short-term reward obtained by applying the SWAP gate.
\begin{equation}\label{equ-reward}
	reward(sw,\pi)= \sum_{g\in L_0} (d(g,\pi)-d(g,sw\cdotp \pi))
\end{equation}

\emph{Step 2.} According to the reward, select the best SWAP gate. We will choose the SWAP gate with the best reward. If there are multiple, we will choose the one through roulette-wheel selection. As shown in (\ref{equ-proba}), the chosen probability of each SWAP gate depends on its fidelity. The higher the fidelity of a SWAP gate, the higher the probability of being selected. The SWAP gate can be realized by three CNOT gates, so its fidelity is the product of the fidelity of these CNOT gates. We show the fidelity of a SWAP gate in (\ref{equ-fid}), where $err()$ represents the error rate of the CNOT gate.

\begin{equation}\label{equ-proba}
	p(sw_i) = \frac{fid(sw_i)}{\sum_{i=1}^{n} fid(sw_i)}
\end{equation}

\begin{equation}\label{equ-fid}
	fid(sw(v_i,v_j)) = (1-err(v_i,v_j))^2 \cdotp (1-err(v_j,v_i))
\end{equation}

\emph{Step 3.} Insert the SWAP gate and update the environment status $(\pi,vcir,pcir)$. First, update $\pi$ by applying the SWAP gate. Second, renew $vcir$ by removing from it all gates executable on $\pi$. Third, renew $pcir$ by inserting the SWAP gate and all just executed gates into it.

The agent will repeat the above decision-making steps until $vcir$ is empty. At last, $pcir$ is the physical circuit that satisfies the hardware constraint. 

Given the coupling graph $CG(V,E)$, in the worst case, we should calculate (\ref{equ-reward}) for $|E|$ candidate SWAP gates. Since the front layer contains $|V|/2$ two-qubit gates at most,  the time complexity of individual agent decision-making is $O(|E| \cdotp |V|/2 )$.

\subsection{Qubit routing based on multi-agent cooperation}\label{sec-multi-routing}

\begin{figure}
 \centering
 \begin{tikzpicture}
 \tikzstyle{every node}=[font=\scriptsize]
 \tikzset{dot/.style = {circle, fill, minimum size=#1, inner sep=0pt, outer sep=0pt},
  dot/.default = 15pt }
 
 \draw[draw=gray, thin] (-0.8,-1.1) rectangle (6.4,2.5);
 \node[dot=80pt, draw=gray, fill=blue!5, thin] at (0.8,0.5) {};
 \node[dot, draw=gray, fill=blue!20, thin] at (0,1) {B};
 \node[dot, draw=gray, fill=blue!20, thin] at (0,0) {};
 \node[dot, draw=gray, fill=blue!20, thin] at (0.8,1) {};
 \node[dot, draw=gray, fill=blue!20, thin] at (0.8,0) {};
 \node[dot, draw=gray, fill=blue!20, thin] at (1.6,1) {};
 \node[dot, draw=gray, fill=blue!20, thin] at (1.6,0) {W};
 \draw [<-,dashed] (0.18,0.8) -- (1.4,0.2);
 
 \draw [<->,dashed] (2.2,0.5) -- (3.4,0.5);
 
 \node[dot=80pt, draw=gray, fill=blue!5, thin] at (4.8,0.5) {};
 \node[dot, draw=gray, fill=green!20, thin] at (4,1) {B};
 \node[dot, draw=gray, fill=green!20, thin] at (4,0) {};
 \node[dot, draw=gray, fill=green!20, thin] at (4.8,1) {};
 \node[dot, draw=gray, fill=green!20, thin] at (4.8,0) {};
 \node[dot, draw=gray, fill=green!20, thin] at (5.6,1) {};
 \node[dot, draw=gray, fill=green!20, thin] at (5.6,0) {W};
 \draw [<-,dashed] (4.18,0.8) -- (5.4,0.2);
 
 \node at (2.8,0.7) {global};
 \node at (2.9,2.2) {Agent Population};
 \node at (0.85,1.55) {Group1};
 \node at (4.85,1.55) {Group2};
 \node at (1.5,0.5) {local};
 \node at (5.5,0.5) {local};
 \end{tikzpicture}
 \label{fig-co}
 \caption{Multi-agent cooperation based on local and global information exchange.}
\end{figure}
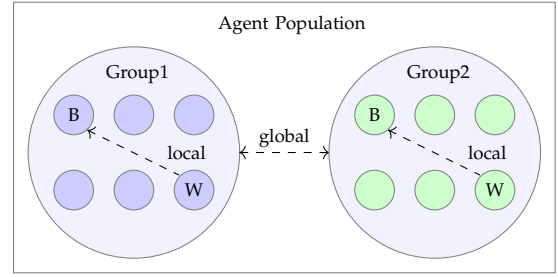

Just a single agent is not enough to achieve essential improvements in circuit quality. To overcome this, we include multiple agents to form a population and adopt an intelligent information-exchanging mechanism. This communication mechanism is similar to that used by the shuffled frog-leaping algorithm (SFLA~\cite{eusuff2006shuffled}). SFLA is a memetic meta-heuristic for combinatorial optimization. It combines the benefits of the local and global search to converge to the global optima with a good speed and likelihood. However, it is not practicable to treat the qubit routing problem as a traditional combinatorial optimization problem and solve it with SFLA. The reason is that,  unless we make multi-stage decisions meticulously, we generally cannot obtain a feasible solution to this problem through mere observation. Even worse, we cannot get another feasible solution by simply perturbing one feasible solution. To adapt this communication mechanism to the qubit routing problem, we take the partial solution generated by each agent as the sharing information rather than the complete solution. Taking Fig.~\ref{fig-co} as an example, agents are arranged into two groups and cooperate through the local and global information exchange. Agents in the same group communicate through the local shared information, making the worst agent (labeled with W) converge to the best one (labeled with B); agents in different groups communicate through the global information exchange. The global and local information exchange, plus the decision-making of every agent, constitute the iteration of our qubit routing approach based on multi-agent cooperation.  As shown in Algorithm~\ref{algo2}, our multi-agent qubit routing approach mainly consists of six steps.

\LinesNumbered
\begin{algorithm}[t]
	\KwIn{Logical circuit $LC(Q,G)$, Coupling graph $CG(V,E)$}
	\KwOut{Physical circuit $PC(Q,G)$}
	$agents = initialize(m,n)$\;
	\While{$is\_done(agents)$} {
		$A=fitness\_rank(agents)$\;
		$S=agent\_partition(A)$\;
		$agents=agent\_evolution(S)$\;
		$agent\_decision(agents)$\;
	}
	\Return $PC$\;
	\caption{Qubit routing based on multi-agent cooperation}
	\label{algo2}
\end{algorithm}

\emph{Step 1.} Initialize all agents. Similar to SFLA, there are $N=m*n$ agents partitioned into $m$ groups. We set the initial state $(\pi,vcir,pcir)$ of the environment for each agent, respectively.  More specifically, $\pi$ is set to be the initial mapping returned by Algorithm~\ref{algo1}; $vcir$ and $pcir$ are updated by checking executable gates in the original circuit based on $\pi$. Algorithm~\ref{algo1} can generate various high-quality initial mappings through multiple runs, ensuring the diversity and quality of the initial population.

\emph{Step 2.} Check termination condition. An agent completes the qubit routing process if the property $vcir$ of its environment is empty. When the number of agents that have completed routing exceeds $n$, we stop this algorithm because the remaining agents have a low probability of producing better physical circuits. Otherwise, go to the next step.

\emph{Step 3.} Rank agents according to fitness (global information exchange). We compute the fitness value for each agent and sort the agents in descending order according to the fitness value, leading to an ordered list of agents $A=[A_1,A_2,\cdots,A_{N-1}]$. The agent fitness is the circuit fidelity estimated based on its environmental state $(\pi,vcir,pcir)$, as shown in (\ref{equ-fit}). In (\ref{equ-fit}), $fid(pcir)$ represents the fidelity of the partial physical circuit $pcir$ and is obtained through (\ref{equ-esp-pcir}); $fid_{worst}$, which is derived from (\ref{equ-esp-worst-pcir}), denotes the worst-case fidelity of the physical circuit corresponding to the remaining logical circuit $vcir$.

\begin{equation}\label{equ-fit}
	fit(agent) = fid(pcir)\cdotp fid_{worst}(vcir)
\end{equation}

The first part of (\ref{equ-esp-pcir}) is a penalty factor, which indicates that the fidelity of the quantum circuit decreases exponentially as the circuit depth increases. The second part of (\ref{equ-esp-pcir}) is the fidelity product of all quantum gates (including the SWAP gates inserted so far) in $pcir$. In (\ref{equ-esp-pcir}), $dep_{pcir}$ is the circuit depth of $pcir$; $|G_{ori}|$ is the number of two-qubit gates in the original logical circuit; $D$ is the diameter of the coupling graph; $|G_{ori}|\cdotp (3D-2)$ denotes the worst-case circuit depth of the final physical circuit; $err()$ is the error rate of the two-qubit gate (CNOT gates in this paper), which comes from IBM's device calibration data. 

\begin{equation}\label{equ-esp-pcir}
	fid(pcir) = e^{\frac{-dep_{pcir}}{|G_{ori}|\cdotp (3D-2)}}\cdotp \prod_{g\in pcir}(1-err(g))
\end{equation}

(\ref{equ-esp-worst-pcir}) estimates the worst-case circuit fidelity of $vcir$, where $e_{max}$ denotes the maximum error rate of all possible two-qubit gates.

\begin{equation}\label{equ-esp-worst-pcir}
	fid_{worst}(vcir) = (1-e_{max})^{|G_{vcir}|\cdotp (3D-2)}
\end{equation}

\begin{figure}[!t]
	\begin{center}
		\begin{tikzpicture}
			\tikzset{dashbox/.style={thin,dashed}}
			\tikzset{solidbox/.style={thin}}
			\draw (0,0) node(s0) {$S^0$};
			\draw (0,-0.8) node(s0a0) {$S^0_0 = A_0$};
			\draw (0,-1.6) node(s0a3) {$S^0_1 = A_3$};
			\draw (0,-2.4) node(s0a6) {$S^0_2 = A_6$};
			\draw (0,-3.2) node(s0a9) {$S^0_3 = A_9$};
			\draw[dashbox] (-1,0.4) rectangle (1,-3.6);
			\draw[solidbox] (-0.9,-0.5) rectangle (0.9,-1.1);
			\draw[solidbox] (-0.9,-1.3) rectangle (0.9,-1.9);
			\draw[solidbox] (-0.9,-2.1) rectangle (0.9,-2.7);
			\draw[solidbox] (-0.9,-2.9) rectangle (0.9,-3.5);
			\draw (2.2,0) node(s1) {$S^1$};
			\draw (2.2,-0.8) node(s1a1) {$S^1_0 = A_1$};
			\draw (2.2,-1.6) node(s1a4) {$S^1_1 = A_4$};
			\draw (2.2,-2.4) node(s1a7) {$S^1_2 = A_7$};
			\draw (2.2,-3.2) node(s1a10) {$S^1_3 = A_{10}$};
			\draw[dashbox] (1.2,0.4) rectangle (3.2,-3.6);
			\draw[solidbox] (1.3,-0.5) rectangle (3.1,-1.1);
			\draw[solidbox] (1.3,-1.3) rectangle (3.1,-1.9);
			\draw[solidbox] (1.3,-2.1) rectangle (3.1,-2.7);
			\draw[solidbox] (1.3,-2.9) rectangle (3.1,-3.5);
			\draw (4.4,0) node(s2) {$S^2$};
			\draw (4.4,-0.8) node(s2a2) {$S^2_0 = A_2$};
			\draw (4.4,-1.6) node(s2a5) {$S^2_1 = A_5$};
			\draw (4.4,-2.4) node(s2a8) {$S^2_2 = A_8$};
			\draw (4.4,-3.2) node(s2a11) {$S^2_3 = A_{11}$};
			\draw[dashbox] (3.4,0.4) rectangle (5.4,-3.6);
			\draw[solidbox] (3.5,-0.5) rectangle (5.3,-1.1);
			\draw[solidbox] (3.5,-1.3) rectangle (5.3,-1.9);
			\draw[solidbox] (3.5,-2.1) rectangle (5.3,-2.7);
			\draw[solidbox] (3.5,-2.9) rectangle (5.3,-3.5);
		\end{tikzpicture}
	\end{center}
	\caption{\label{fig-group} Partition of agents.}
\end{figure}
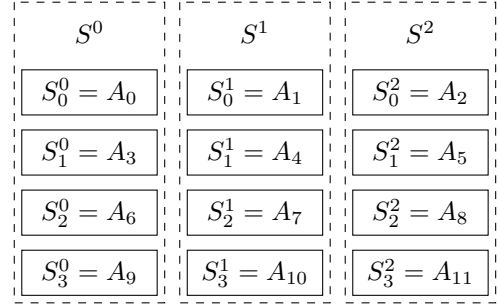

\emph{Step 4.} Partition agents in groups. We divide the sorted list $A$ into $m$ groups ($S^0,S^1,\cdots,S^{m-1}$) so that each group has $n$ agents. We assign the agents in $A$ to different groups according to (\ref{equ-group}). An example of agent grouping is shown in Fig.~\ref{fig-group}, which divides 12 ordered agents into three groups.

\begin{equation}\label{equ-group}
\begin{split}
    S^k = [S^k_j = A_{k+m(j-1)},j=0,1,\dots,n-1] \\ where \quad k=0,1,\dots,m-1.
\end{split}
\end{equation}

\emph{Step 5.} Evolve the worst agent in each group (local information exchange). For the agent group $S^k\,(k\in[0,m-1])$, the agent with the worst fitness is $S^k_{n-1}$, while the agent with the best fitness is $S^k_0$. We decide whether to update $S^k_{n-1}$ according to (\ref{equ-condi}). In (\ref{equ-condi}), $r$ is a random number in [0,1), $\Delta G_{vcir}$ denotes the difference of $S^k_{n-1}$ and $S^k_0$ in terms of the gate count of their $vcir$ property, and $C$ is a predefined constant. (\ref{equ-condi}) shows that the greater the difference between the best agent and the worst agent in the gate count or the fidelity, the higher the probability of evolving the worst agent. It is worth noting that we introduce the random number $r$ in (\ref{equ-condi}) to prevent the worst agent in each group from converging on the best agent (especially the best agent without a considerable advantage) too readily, thus ensuring the diversity of the agent population. 

\begin{equation}\label{equ-condi}
	r\cdotp \Delta G_{vcir} \cdotp \frac{fit(S^k_0)}{fit(S^k_{n-1})} > C
\end{equation}

\begin{equation}\label{equ-condi2}
	S^k_{n-1} = S^k_0
\end{equation}

\emph{Step 6.} Decision-making of agents. As discussed in the last section, each agent independently decides which SWAP gate to insert and updates its environment by applying the chosen SWAP gate. Then, go to Step 2.

As shown above, different from solving the traditional combinatorial optimization problem, we continuously update the partial solution rather than the complete solution for each agent through multiple iterations. When Algorithm~\ref{algo2} stops its loop, it will return the physical circuit corresponding to the agent with the best fitness.

\subsection{Time complexity}

In the worst case, an agent makes approximately $O(|G|\cdotp D)$ decisions to complete the qubit routing task.  In addition, an agent makes only one decision (in Step 6) in each loop. Therefore,  Algorithm~\ref{algo2} has at most $O(|G|\cdotp D)$ loops.  In each loop, the decision-making and sorting of $N$ agents take $O(N\cdotp |E| \cdotp |V|/2 )$  and $O(N\cdotp \log N)$ time, respectively. Since $|E|\leq |V|(|V|-1)/2$, $D<|V|$, and $\log N$ is relatively small, the worst-case time complexity of Algorithm~\ref{algo2} is roughly $O(N \cdotp |G|\cdotp |V|^4)$. Table~\ref{tab-time-complexity} gives the time complexity of several qubit routing approaches used in the experimental evaluation. As shown, the time complexity of our approach is polynomial in all parameters, so is HA's, whereas DL's time increases exponentially as the search depth $k$ grows. In addition, the time 
required by our algorithm is linear with the number of agents ($N$).

\begin{table}[!t]
    \centering
	\caption{Time complexity of routing approaches}
    \setlength\extrarowheight{4.6pt}
    \begin{tabular}{c|c}
        \hline
        \multicolumn{1}{c|}{ Routing algorithm} & \multicolumn{1}{c}{Time complexity} \\ \hline
        HA~\cite{niu2020hardware} & $O(|G|^2 \cdotp |V|^3)$ \\
        DL~\cite{li2020qubit} & $O(|G|^2 \cdotp |V|^{2k})$  \\
        Ours & $O(N \cdotp |G|\cdotp |V|^4)$  \\ \hline   
    \end{tabular}
    \label{tab-time-complexity}
\end{table}

\section{Experimental evaluation}\label{sec-experiment}

In this section, to evaluate the effectiveness in improving circuit fidelity, we compare our multi-agent methodology with the state-of-the-art approach.

\subsection{Experimental configuration}

\subsubsection{Algorithms for comparison}
We compare the variation-aware multi-agent approach (MA) with the other three algorithms. First, to prove the effectiveness of the error-variation-aware strategy, we implement another multi-agent approach which is variation-agnostic (MA\_NA). MA\_NA is almost the same as MA, but it always assumes that all gates have the same error rate when calculating all fidelity-related equations (such as (\ref{equ-fid}) and (\ref{equ-condi})). Second, to understand how much improvement MA can achieve in the circuit fidelity,  we compare it with another variation-aware method (HA~\cite{niu2020hardware}) in the success rate of physical circuits. HA also aims to improve the overall circuit fidelity and show advantages over two approaches~\cite{murali2019noise,li2019tackling} integrated into QISKIT. To obtain the experiment data of HA on the latest heavy-hex architecture, we download its code from Github and evaluate it with the same experimental configuration as our approach. Third, although the number of gates is not our first optimization goal, we also evaluate the gate count inserted by MA. To this end, we compare MA with the approach (DL~\cite{li2020qubit}) that shows considerable improvements in reducing gates.  

\subsubsection{Target quantum devices}
We use two well-known quantum devices of IBM as the target hardware platform. The first is a 16-qubit quantum processor, IBMQ\_guadalupe. The topology of IBMQ\_guadalupe is based on the latest heavy-hex lattice, which can provide a better Quantum Volume. Besides IBMQ\_guadalupe, we also take another 20-qubit quantum processor, IBMQ\_tokyo. Although BMQ\_tokyo has already been deprecated due to high qubit frequency collision and high gate error rate, the experimental data of DL is given based on it. Therefore, we choose this architecture to compare with DL. 
 
\subsubsection{Benchmark circuit} 
We collect various benchmark circuits from the previous work~\cite{li2020qubit,li2019tackling,zulehner2018efficient}. For easy illustration, we divide all the benchmarks into three classes. The small-scale class refers to benchmarks with less than 100 gates; the medium-scale to those with 100 to 1,000 gates; and the large-scale to those with more than 1,000 gates.

\subsubsection{Simulator of real quantum device}
We use the simulator to execute quantum circuits rather than real quantum devices for three reasons. First, publicly available resources of quantum hardware are very scarce. Taking IBM as an example, they only provide public access to several 5-qubit devices. Such devices with very few qubits are not enough for demonstrating the advantage of automated approaches for quantum circuit mapping. Moreover, current quantum devices only allow very small-scale quantum circuits due to high gate error and low coherence time. In contrast, the simulator can support much larger circuits by adjusting the gate error or ignoring some noise sources. Therefore, experimenting on a simulator is more conducive to fully showing the performance differences of various algorithms.

We use the Qiskit Aer~\cite{wille2019ibm} to generate a noise model for a hardware device. This model takes the hardware calibration information as the input and thus can approximate the errors occurring on the actual device. We set two versions of noise models, that is, the basic one and the improved one. The basic model comes from the 16-qubit device IBMQ\_guadalupe, while the improved model is obtained by reducing the gate error of the basic model by 10x. The 10x reduced gate error reaches the threshold of quantum error correction, so the improved model is a reasonable approximation of future quantum devices. In our experiment, we create a simulator based on the basic noise model for the small-scale benchmarks. Moreover, we also implement a  simulator based on the improved noise model for the other circuits with hundreds of gates. However, this improved model alone is still insufficient to allow large circuits to output reliable results due to their long circuit depth. Therefore, the simulator for large circuits neglects the decoherence error. Furthermore, to focus on the impact of gate errors on circuit fidelity, both simulators also ignore the readout error, which happen at most once for each qubit during computation.

\subsubsection{Figures of merit}
Our approach aims to improve circuit fidelity, so the first merit is the probability of success trials (PST). As shown in (\ref{equ-pst}), PST gives the success rate of running the physical circuit on the hardware. In the experiment, we run each physical circuit 8192 times. In addition, we also consider the number of gates and the circuit depth, another two important metrics of circuit quality.
\begin{equation}\label{equ-pst}
	pst = \frac{\#suc\_trail}{\#total\_trail}
\end{equation}

\subsubsection{Runtime environment}
We implement our algorithm in Python and conduct all experiments on a personal computer with i7-4710HQ CPU and 16GB memory. In all experiments, MA and MA\_NA both have 100 agents partitioned into 20 groups, and the constant $C$ in (\ref{equ-condi}) is set to 1. We choose the experimental configuration empirically to achieve a good trade-off of the circuit quality and the time. Our approach is publicly available on Github.

\subsection{Effectiveness of error variation-aware strategy }
\begin{figure}[!t]
	\centering
	\includegraphics[scale=0.38]{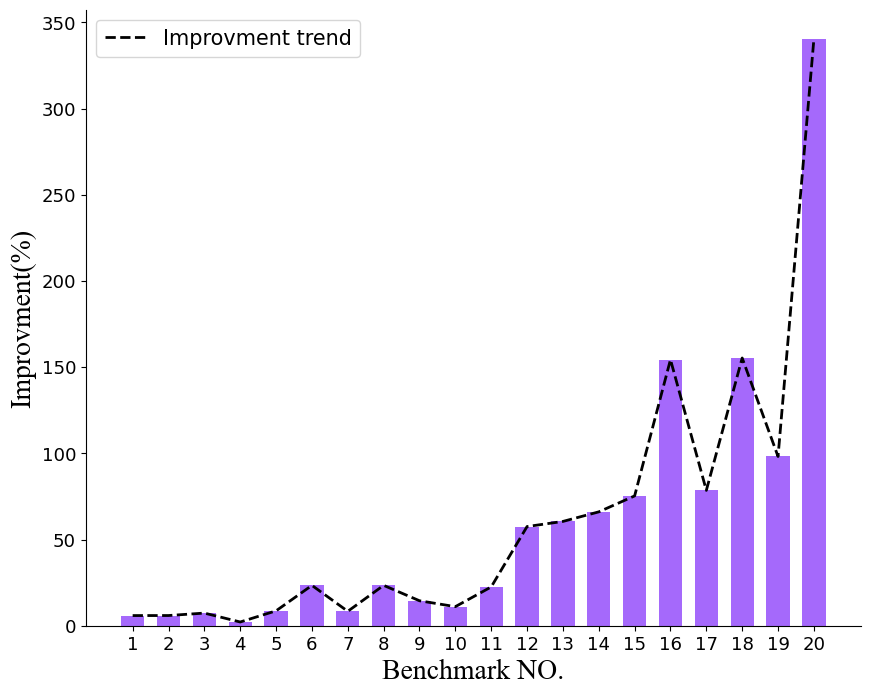}
	\caption{Comparison of MA\_NA and MA}
	\label{fig-Comp-MA-NA}
\end{figure}

To verify the effectiveness of the error variation-aware strategy in improving circuit fidelity, we use IBMQ\_guadalupe as the target architecture and compare MA with MA\_NA on various benchmarks. MA\_NA assumes every two-qubit gate has the same error rate, so it gives preference to reducing the number of gates rather than improving the circuit fidelity. We can see from Table~\ref{tab-Comp-MA-NA} that the number of additional gates required by MA\_NA is less than or equal to that of MA. Although MA\_NA has advantages in reducing gates, its circuit fidelity is much worse than MA. On all benchmarks, MA improves PST by 60.95\% on average and 340.28\% at most. Moreover, the improvement becomes more significant with the increase of the circuit scale, as shown in Fig.~\ref{fig-Comp-MA-NA}.  Given that MA and MA\_NA use the same multi-agent scheme, we attribute this improvement of MA to the error variation-aware strategy.

We also use MA and MA\_NA to experiment on large-scale circuits. However, even with the simulator based on the improved noise model, the PST value is still too low, indicating the output is unreliable and meaningless. It confirms that quantum error correction is essential for the reliable execution of large-scale quantum circuits. 

\begin{table*}[!t]
	\centering
	\caption{Comparison of MA\_NA and MA}
	\begin{tabular}{clcc|cc|cc|c}
		\hline
		\multicolumn{4}{c|}{Benchmark circuits}  & \multicolumn{2}{c|}{MA\_NA}  & \multicolumn{2}{c|}{MA}  &  Imprv \\ 
		\multicolumn{1}{c}{$NO.$} & {$name$} & $n$ & $g_{ori}$ & $g_{add}$ & $pst$ & $g_{add}$ & $pst$ & $p_{imp}$ \\ \hline
		1 & graycode6\_47 & 6 & 5 & 0 & 95.70\% & 0 & 98.73\% & 5.89\%  \\ 
		2 & xor5\_254 & 6 &	7 &	6 &	87.89\%	& 6	& 93.07\% &	5.91\% \\ 
		3 & 4mod5-v1\_22	&	5	&	21	&	9	&	85.94\%	&	9	&	85.74\%	&	7.33\%	\\	
		4 & mod5mils\_65	&	5	&	35	&	18	&	64.45\%	&	18	&	69.34\%	&	2.16\%	\\	\
		5 & alu-v0\_27	&	5	&	36	&	21	&	71.19\%	&	21	&	77.34\%	&	8.64\%	\\	
		6 & decod24-v2\_43	&	4	&	52	&	27	&	48.83\%	&	27	&	60.25\%	&	23.40\%	\\	
		7 & mod5d2\_64	&	5	&	53	&	33	&	54.88\%	&	33	&	59.47\%	&	8.36\%	\\	
		8 & 4gt13\_92	&	5	&	66	&	39	&	53.81\%	&	39	&	66.41\%	&	23.41\%	\\	
		9 & alu-v0\_26	&	5	&	84	&	54	&	49.90\%	&	54	&	57.13\%	&	14.48\%	\\	
		10 & 4gt5\_76	&	5	&	91	&	57	&	45.90\%	&	57	&	50.98\%	&	11.06\%	\\	
		11 & qft\_10	&	10	&	200	&	96	&	67.29\%	&	96	&	82.42\%	&	22.50\%	\\	
		12 & 4gt4-v0\_72	&	6	&	258	&	171	&	48.05\%	&	171	&	75.68\%	&	57.52\%	\\	
		13 & sym6\_316	&	14	&	270	&	207	&	43.75\%	&	219	&	70.21\%	&	60.49\%	\\	
		14 & rd53\_135	&	7	&	296	&	216	&	41.41\%	&	219	&	68.75\%	&	66.04\%	\\	
		15 & mod8-10\_177	&	6	&	395	&	306	&	32.03\%	&	318	&	56.15\%	&	75.30\%	\\	
		16 & cnt3-5\_180	&	16	&	485	&	348	&	21.39\%	&	360	&	54.39\%	&	154.34\%	\\	
		17 & qft\_16	&	16	&	512	&	288	&	32.71\%	&	303	&	58.40\%	&	78.51\%	\\	
		18 & rd53\_133	&	7	&	580	&	387	&	19.43\%	&	423	&	49.61\%	&	155.28\%	\\	
		19 & 4gt4-v0\_73	&	6	&	635	&	261	&	31.84\%	&	267	&	63.09\%	&	98.16\%	\\	
		20 & con1\_216	&	9	&	954	&	666	&	7.03\%	&	696	&	30.96\%	&	340.28\%	\\	\hline
	\end{tabular}
	\label{tab-Comp-MA-NA}
	
	$name$: benchmark circuit; $n$: the number of qubits; $g_{ori}$: the number of gates in the original circuit; $g_{add}$: the number of additional CNOT gates inserted; $pst$: the success rate; $P_{imp}$: the improvement rate of MA in $pst$ compared to MA\_NA.
\end{table*}

\subsection{Evaluation of circuit fidelity}
\begin{figure}[!t]
	\centering
	\includegraphics[scale=0.38]{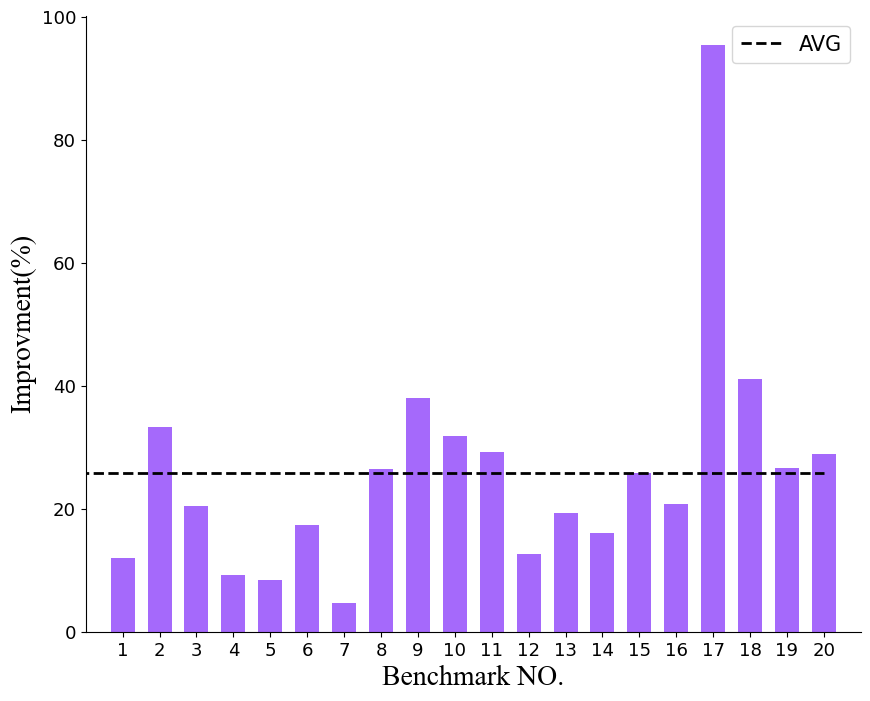}
	\caption{Comparison of MA and HA}
	\label{fig-Comp-MA-HA}
\end{figure}

Both HA and MA consider the gate-error variation in the quantum circuit mapping process.  We use IBMQ\_guadalupe as the target platform to evaluate the efficiency of these two approaches in improving circuit fidelity. We show the experiment results in Table~\ref{tab-Comp-MA-HA} and Fig.~\ref{fig-Comp-MA-HA}. Compared with HA, MA improves PST by 25.86\% on average and 95.42\% at most. Even on several benchmarks (mod5d2\_64 and 4gt4-v0\_72), MA requires more gates or circuit depth than HA, but it still shows better circuit fidelity. We owe the advantage of MA in improving the overall circuit quality to the multi-agent scheme. HA is a typical approach of quantum circuit mapping. It performs the locally optimal action in each step while discarding other possibilities. In contrast, MA updates the environment state of the agent with the worst fitness in each group while retaining the status of other agents. Combining with the local search (Step 5 of Algorithm~\ref{algo2})and global search (Step 3 of Algorithm~\ref{algo2}), MA can evolve the overall circuit quality of the agent population gradually as the iteration goes. 

Both HA and MA have polynomial time complexity, so the time overhead of both approaches is small for all benchmarks considered. In addition, since MA contains multiple agents, the time it consumes is generally more than that of HA.

\begin{table*}[!t]
	\centering
	\caption{Comparison of MA and HA}
	\begin{tabular}{clcc|cccc|cccc|c}
		\hline
		\multicolumn{4}{c|}{Benchmark circuits}  & \multicolumn{4}{c|}{HA}  & \multicolumn{4}{c|}{MA}  &  Imprv \\ 
		\multicolumn{1}{c}{$NO.$} & {$name$} & $n$ & $g_{ori}$ & $g_{add}$ & $dep$ & $pst$ & $t$ & $g_{add}$ & $dep$ & $pst$ & t & $p_{imp}$ \\ \hline
		1 & graycode6\_47 & 6 & 5 & 15 & 16  & 88.18\% & 0.01 & 0 & 5 & 98.73\%  & 0.01  & 11.96\% \\ 
		2 & xor5\_254 & 6 & 7 & 27 & 24 & 69.82\% & 0.02 & 6 & 10 & 93.07\% & 0.03 & 33.29\% \\ 
		3 & 4mod5-v1\_22 & 5 & 21 &	36 & 47 & 71.19\% &	0.01 & 9 & 23 &	85.74\% & 0.05 & 20.44\% \\ 
		4 & mod5mils\_65 &	5 &	35 & 21 & 40 & 63.48\% & 0.02 &	18 & 44 & 69.34\% &	0.11 & 9.23\%  \\ 
		5 & alu-v0\_27 & 5 & 36 & 24 & 51 &	71.39\% & 0.02 & 21 & 41 & 77.34\% & 0.13 &	8.34\%  \\ 
		6 & decod24-v2\_43 & 4 & 52 & 33 & 61 &	51.37\% & 0.02 & 27 & 63 &	60.25\% & 0.16 & 17.30\%  \\ 
		7 & mod5d2\_64 & 5 & 53 & 33 & 62 &	56.84\% & 0.02 & 33 & 71 & 59.47\% & 0.18 &	4.64\%  \\ 
		8 & 4gt13\_92 &	5 &	66 & 63 & 98 & 52.54\% & 0.03 &	39 & 82 & 66.41\% &	0.22 & 26.39\%  \\ 
		9 & alu-v0\_26 & 5 & 84 & 66 & 117 & 41.41\% & 0.03 & 54 & 113 & 57.13\% & 0.28 & 37.97\%  \\ 
		10 & 4gt5\_76 & 5 & 91 &	96 & 143 & 38.67\% & 0.03 &	57 & 123 & 50.98\% & 0.30 &	31.82\% \\ 
		11 & qft\_10 & 10 & 200 & 294 & 314 & 63.77\% & 0.14 & 96 & 150 & 82.42\% & 1.01 & 29.25\%  \\ 
		12 & 4gt4-v0\_72 & 6 & 258 & 189 & 315 & 	67.19\% & 	0.06 & 	171 & 	317 & 	75.68\%	 & 0.96  & 	12.65\%  \\ 
		13 & sym6\_316 & 14	 & 270 & 	348 & 	373 & 	58.89\% & 	0.20 & 	219 & 	315 & 	70.21\% & 	1.26  & 	19.24\%  \\ 
		14 & rd53\_135 & 	7 & 	296	 & 252 & 	391 & 	59.28\% & 	0.11 & 	219	 & 386 & 	68.75\% & 	1.18  & 	15.98\%  \\ 
		15 & mod8-10\_177 & 	6 & 	395 & 	432	 & 648 & 	44.63\% & 	0.16 & 	318 & 	593 & 	56.15\% & 	1.55  & 	25.82\%  \\ 
		16 & cnt3-5\_180	 & 16 & 	485 & 	468 & 	555	 & 45.02\% & 	0.20 & 	360	 & 555 & 	54.39\% & 	2.21  & 	20.82\%  \\ 
		17 & qft\_16	 & 16 & 	512	 & 864 & 	881 & 	29.88\%	 & 0.38 & 	303 & 	298 & 	58.40\%	 & 3.60  & 	95.42\%    \\ 
		18 & rd53\_133 & 	7 & 	580	 & 528 & 	801 & 	35.16\% & 	0.21 & 	423	 & 788 & 	49.61\% & 	2.48  & 	41.11\%  \\ 
		19 & 4gt4-v0\_73 & 	6 & 	635 & 	345 & 	541	 & 49.80\% & 	0.11 & 	267 & 	524 & 	63.09\% & 	1.65  & 	26.67\% \\ 
		20 & con1\_216 & 	9 & 	954 & 	801 & 	1239 & 	24.02\%	 & 0.39	 & 696 & 	1234 & 	30.96\% & 	3.86  & 	28.86\%  \\ \hline	
	\end{tabular}
	\label{tab-Comp-MA-HA}

	$name$: benchmark circuit; $n$: the number of qubits; $g_{ori}$: the number of gates in the original circuit; $g_{add}$: the number of auxiliary CNOT gates inserted; $dep$: the depth of the circuit; $pst$: the success rate; $t$: the running time in seconds; $p_{imp}$: the improvement rate of MA in $pst$ compared to HA.
\end{table*}

\subsection{Evaluation of gate overhead}

DL does not consider the gate error issue and aims to minimize the number of gates. DL's experiments use IBMQ\_tokyo as the target platform. To compare with DL, we also give the experiment results of MA with IBMQ\_tokyo as the target architecture, as shown in Table~\ref{tab-Comp-MA-DL}. Although MA takes circuit fidelity as the optimization goal, it can generally reduce the gate overhead. MA achieves improvement in gate count 16.81\% on average and 66.67\% at most. The advantage of MA in gate overhead comes from two main reasons. First, the fitness function of the agent implicitly considers the gate overhead. The fitness decays as the number of gates increases. Second, the multi-agent scheme can effectively improve the overall fidelity of the circuit, and circuit fidelity is a comprehensive metric involving various factors, such as the number of gates, the circuit depth, and the gate error.

\begin{table*}[!t]
	\centering
	\caption{Comparison of MA and DL}
	\begin{tabular}{lcc|cc|cc|c}
		\hline
		\multicolumn{3}{c|}{Benchmark circuits}  & \multicolumn{2}{c|}{DL}  & \multicolumn{2}{c|}{MA}  &  Imprv  \\ 
		\multicolumn{1}{c}{$name$} & $n$ & $g_{ori}$ & $g_{add}$ & $t$ & $g_{add}$ & $t$ & $g_{imp}$  \\ \hline
		4mod5-v1\_22 & 5 & 21 & 0 & 0 & 0 & 0 & 0.00\%  \\ 
		mod5mils\_65	&	5	&	35	&	0	&	0	&	0	&	0.02 	&	0.00\%	\\	
		alu-v0\_27	&	5	&	36	&	9	&	0.07	&	3	&	0.03 	&	66.67\%	\\	
		decod24-v2\_43	&	4	&	52	&	0	&	0	&	0	&	0.02 	&	0.00\%	\\	
		4gt13\_92	&	5	&	66	&	0	&	0	&	0	&	0.01 	&	0.00\%	\\	
		ising\_model\_10	&	10	&	480	&	0	&	0	&	0	&	0.02 	&	0.00\%	\\	
		ising\_model\_13	&	13	&	633	&	0	&	0	&	0	&	0.10 	&	0.00\%	\\	
		ising\_model\_16	&	16	&	786	&	0	&	0	&	0	&	0.21 	&	0.00\%	\\	
		qft\_10	&	10	&	200	&	39	&	1.6	&	30	&	0.61 	&	23.08\%	\\	
		qft\_16	&	16	&	512	&	153	&	21.06	&	129	&	2.96 	&	15.69\%	\\	
		rd84\_142	&	15	&	343	&	72	&	5.01	&	81	&	1.43 	&	-12.50\%	\\	
		adr4\_197	&	13	&	3439	&	630	&	19.85	&	498	&	6.81 	&	20.95\%	\\	
		radd\_250	&	13	&	3213	&	555	&	28.21	&	528	&	6.47 	&	4.86\%	\\	
		z4\_268	&	11	&	3073	&	630	&	16.19	&	372	&	5.85 	&	40.95\%	\\	
		sym6\_145	&	7	&	3888	&	513	&	2.93	&	402	&	5.44 	&	21.64\%	\\	
		misex1\_241	&	15	&	4813	&	786	&	24.62	&	450	&	5.97 	&	42.75\%	\\	
		rd73\_252	&	10	&	5321	&	1095	&	9.72	&	687	&	8.51 	&	37.26\%	\\	
		cycle10\_2\_110	&	12	&	6050	&	1194	&	10.93	&	897	&	12.16 	&	24.87\%	\\	
		square\_root\_7	&	15	&	7630	&	1338	&	228.29	&	834	&	12.92 	&	37.67\%	\\	
		sqn\_258	&	10	&	10223	&	1578	&	17.73	&	1359	&	17.78 	&	13.88\%	\\	
		rd84\_253	&	12	&	13658	&	2352	&	54.35	&	2088	&	29.30 	&	11.22\%	\\	
		co14\_215	&	15	&	17936	&	4257	&	128.33	&	3129	&	55.26 	&	26.50\%	\\	
		sym9\_193	&	11	&	34881	&	5589	&	70.29	&	4971	&	124.83 	&	11.06\%	\\	\hline
	\end{tabular}
	\label{tab-Comp-MA-DL}

	$name$: benchmark circuit; $n$: the number of qubits; $g_{ori}$: the number of gates in the original circuit; $g_{add}$: the number of CNOT gates inserted; $t$: the running time in seconds; $g_{imp}$: the improvement rate of MA in $g_{add}$ compared to DL.
\end{table*}

\section{Discussion}\label{sec-discussion}

As confirmed by the experiment, our multi-agent approach outperforms several noted methods in improving the circuit fidelity. The good performance of our approach mainly originates from the multi-agent cooperation mechanism. In each iteration, our approach discards the partial solution of the worst agent in each group but preserves the others'. In contrast, other approaches always keep a local optimal partial solution while ignoring other possibilities. Therefore, compared with other algorithms, our algorithm is less likely to fall into the local optimum. Moreover, although our approach aims to optimize the circuit fidelity, it can adapt to other optimization objectives (such as the number of gates and circuit depth) by adjusting the reward function and fitness function. Table~\ref{tab-Comp-MA-DL} gives the preliminary evidence that this method can also achieve improvement in the number of gates.

Our approach contains multiple agents. Its time overhead increases linearly with the growth of the number of agents. As a result, our approach generally takes more time than other approaches, which is a disadvantage of our approach. However, since each agent makes its decision independently according to the environment state, we can implement the decision-making step (Step 6 of Algorithm~\ref{algo2}) by parallel programming, so that all agents can make decisions simultaneously. In this way, we can solve this time issue. 

In the decision-making step of our approach, we only consider the SWAP gates that reduce the physical distance for at least one front gate while ignoring other SWAP gates. Although those gates not considered may lead to better physical circuits, involving them in the decision-making process will seriously deteriorate the convergence of our approach, so we get rid of them. One of our future work is to consider more possible SWAP gates while ensuring the convergence of our approach.

There are three key parameters in our multi-agent approach. They are the constant value $C$ in (\ref{equ-condi}), the number $m$ of agent groups, and the number $n$ of agents in each group, respectively. The assignment of parameter $C$ has a significant impact on the quality of the final physical circuit generated. If setting $C$ to 0, the worst individual in the population always evolves into the best individual, so that the whole population is filled with the best individual within a short time, resulting in the loss of population diversity and thus missing the possibility of exploring more potential solutions. Otherwise, if setting C to a large value, it is scarcely likely for individuals to evolve, resulting in that the multi-agent cooperation scheme degenerates into the synchronous mode in which multiple agents complete the qubit routing task independently. Similarly, the two parameters, $m$ and $n$, have considerable influence on the convergence rate and the solution quality. Simply increasing $m$ and $n$ generally slows down the convergence rate but does not necessarily improve the solution quality. We observe from parameter-tuning experiments that when mapping various circuits to a specific quantum architecture, the optimal choice of these three parameters is generally not constant. However, for each parameter, there is a value interval that can bring high-quality solutions. In all the experiments of algorithm evaluation, we set each parameter approximately to the average of its value interval. If adjusting the parameters separately for each experiment, more improvement can be obtained. Therefore, compared with other heuristic algorithms, our approach requires an additional job of parameter adjusting, which is also a drawback.

\section{Conclusion}\label{sec-conclusion}

In this paper, we have proposed a quantum circuit mapping methodology based on multi-agent cooperation. This approach considers the variation of gate error on current quantum hardware and aims to improve the overall circuit fidelity. It involves a population of multiple agents, which are arranged into various groups. Each agent starts from an initial environment state and evolves to the final state by making decisions stepwise. In each decision-making step, all agents exchange information according to a mechanism similar to SFLA, which integrates the local search and the global search. Based on the shared information, the agent with the worst fitness in each group can get a chance to evolve to the best-fitness agent in the group. After numerous iterations, this approach outputs the physical circuit of the agent with the best fitness. The experimental results confirm that this multi-agent scheme can effectively improve the overall circuit fidelity.

\ifCLASSOPTIONcaptionsoff
  \newpage
\fi

\bibliographystyle{IEEEtran}
\bibliography{IEEEabrv,ref}

\begin{IEEEbiography}[{\includegraphics[width=1in,height=1.25in,clip,keepaspectratio]{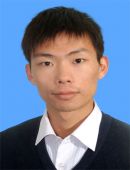}}]{Pengcheng Zhu (S'19)}
received the B.E. and M.E. degrees in computer science from Liaoning Technical University, Fuxin, China, in 2004 and 2007, respectively. He is currently pursuing the Ph.D. degree with the Department of Information and Communication Engineering, Nantong University, China.

He is currently an Associate Professor with the Department of Artificial Intelligence, Suqian University, China. His current research interests include reversible logic design, quantum logic design, and computer-aided design of integrated circuits and systems.
\end{IEEEbiography}

\begin{IEEEbiography}[{\includegraphics[width=1in,height=1.25in,clip,keepaspectratio]{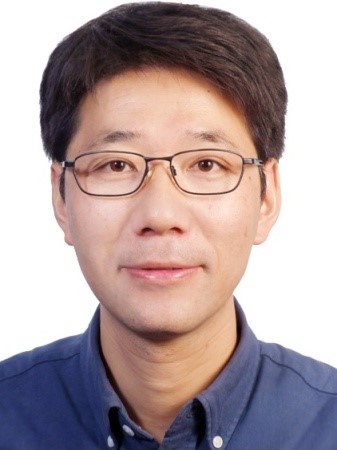}}]{Weiping Ding (M’16-SM’19)}
received the Ph.D. degree in Computer Science, Nanjing University of Aeronautics and Astronautics (NUAA), Nanjing, China, in 2013. In 2016, He was a Visiting Scholar at National University of Singapore (NUS), Singapore. From 2017 to 2018, he was a Visiting Professor at University of Technology Sydney (UTS), Ultimo, NSW, Australia.   

He is currently a Professor with the School of Information Science and Technology, Nantong University, China. His research interests include deep neural networks, multimodal machine learning, granular data mining, and medical images analysis. He served/serves as an Associate Editor of IEEE Transactions on Neural Network and Learning System, IEEE Transactions on Fuzzy Systems, IEEE/CAA Journal of Automatica Sinica, Information Sciences, Neurocomputing, Swarm and Evolutionary Computation. 
\end{IEEEbiography}

\begin{IEEEbiography}[{\includegraphics[width=1in,height=1.25in,clip,keepaspectratio]{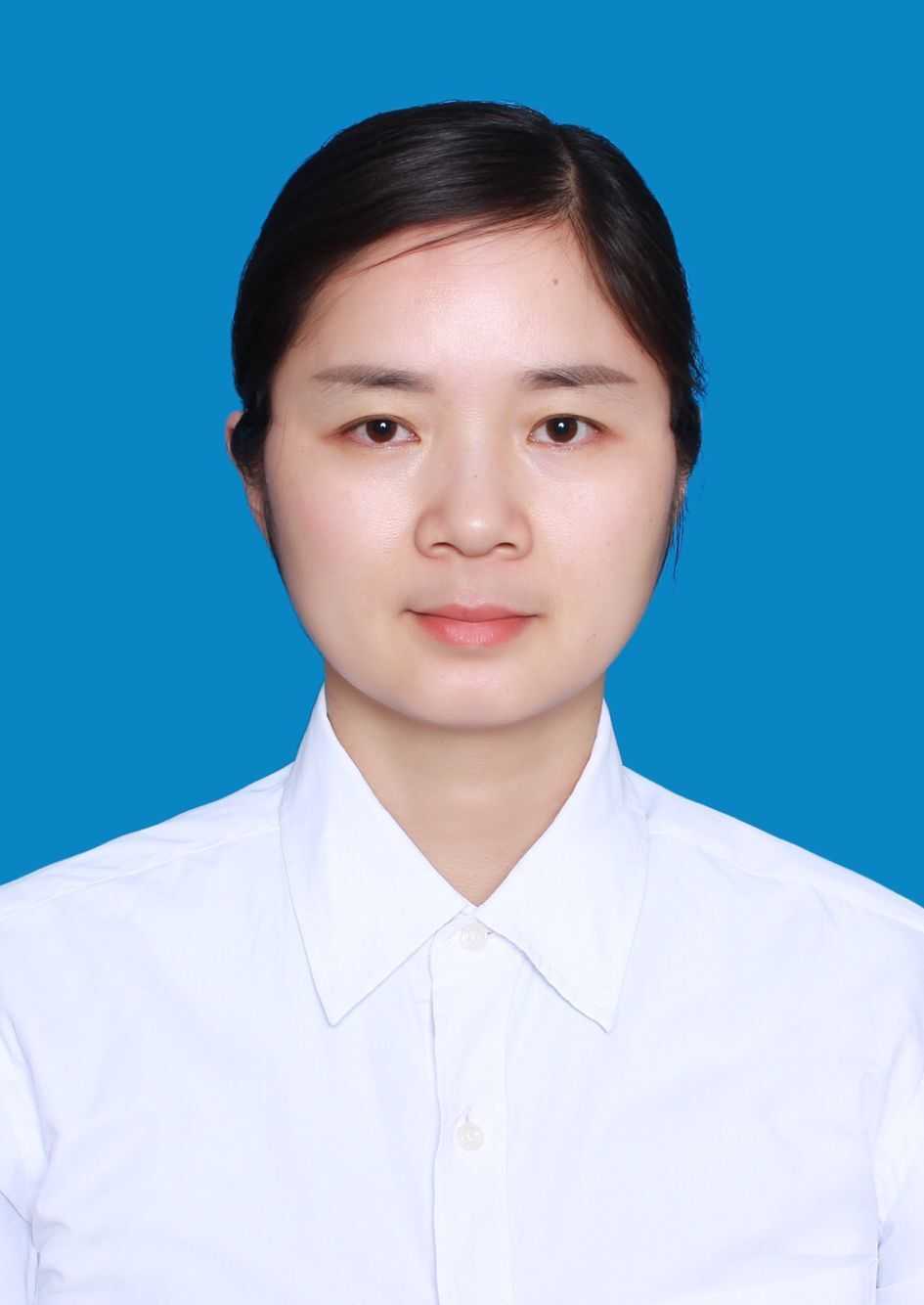}}]{Lihua Wei}
received the B.E. degrees in information and computation Science from Yancheng Teachers University, Yancheng, China, in 2006 and M.E. degrees in computer science from Jiangsu University, Zhenjiang, China, in 2011.  

She is currently a Lecturer with the Department of Information  and Computing Science, Suqian University, China. Her current research interests include reversible logic design, quantum logic design, and computer-aided design of integrated circuits and systems.
\end{IEEEbiography}

\begin{IEEEbiography}[{\includegraphics[width=1in,height=1.25in,clip,keepaspectratio]{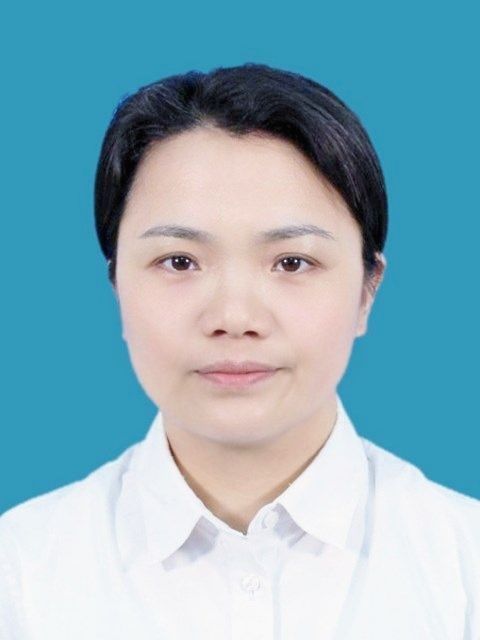}}]{Xueyun Cheng}
 received the B.S. degree in computer science from Nanjing Normal University, Nanjing, China in 2001, and the M.S. degree in computational mathematics from Nanjing Normal University in 2007. 

She is currently an Associate Professor with the School of Information Science and Technology, Nantong University, China. Her current research interests include reversible logic synthesis, quantum information and quantum computation.
\end{IEEEbiography}

\begin{IEEEbiography}[{\includegraphics[width=1in,height=1.25in,clip,keepaspectratio]{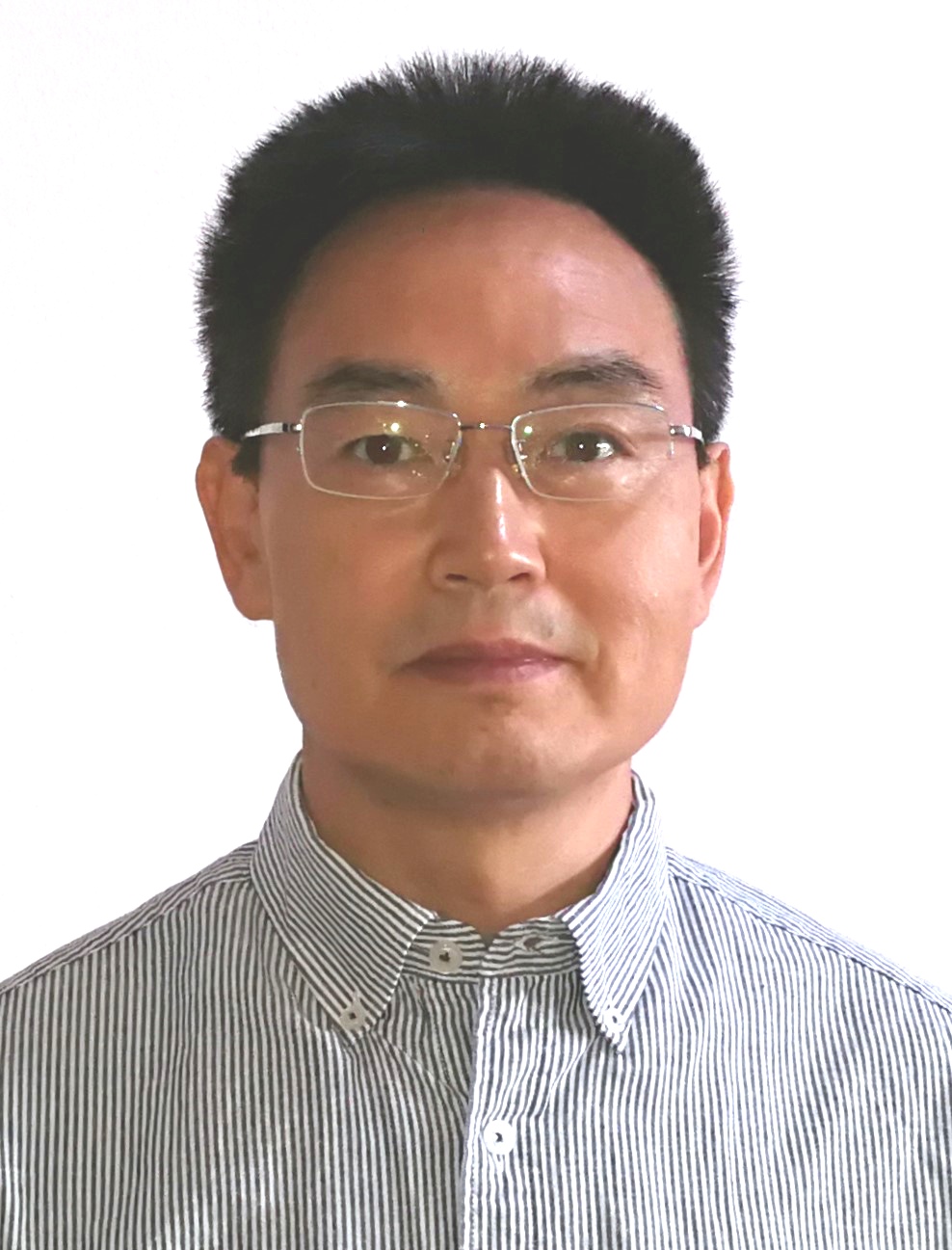}}]{Zhijin Guan}
received the B.S. degree in mathematics from Harbin Normal University, China, in 1986, and the Ph.D. degree in Computer Application Technology from Nanjing University of Aeronautics and Astronautics, China, in 2005. 

He is currently a Professor with the School of Information Science and Technology, Nantong University, China. His current research interests include quantum circuit logic design, secure computing, information-aware and computer-aided intelligent manufacturing, and computer-aided design of integrated circuits and systems.
\end{IEEEbiography}

\begin{IEEEbiography}[{\includegraphics[width=1in,height=1.25in,clip,keepaspectratio]{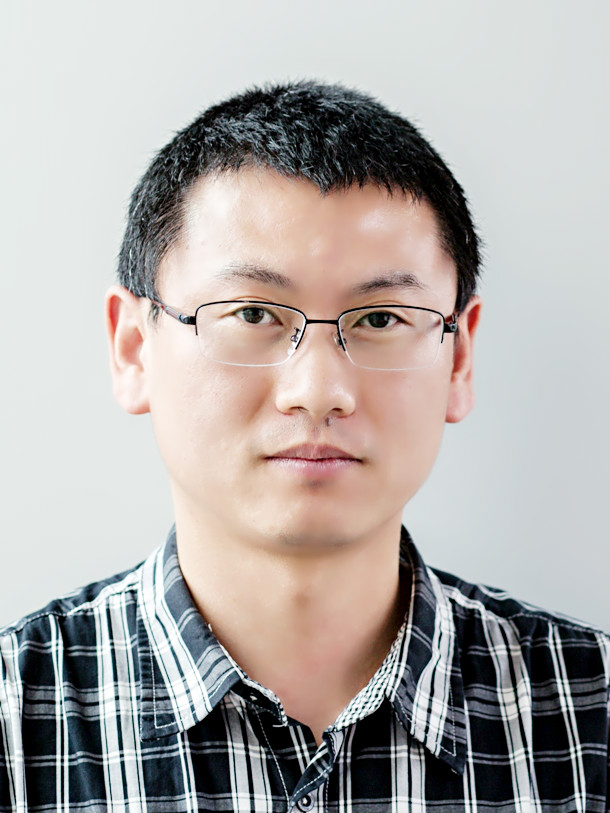}}]{Shiguang Feng}
received the B.S. degree in computer science and technology from Shandong Agricultural University, China, in 2006, and the Ph.D. degree in logic from Sun Yat-sen University, China, in 2012, and the Doctor of Natural Science degree in computer science from Leipzig University, Germany, in 2016.

He is currently an Associated Professor with the School of Information Science and Technology, Nantong University, China. His current research interests include complexity theory and mathematical logic.
\end{IEEEbiography}




\end{document}